\documentclass[12pt]{article}

\usepackage[margin=1in]{geometry}
\usepackage{amsmath, amsfonts}
\usepackage{enumerate}
\usepackage{graphicx}
\usepackage{titling}
\usepackage{url}
\usepackage{xcolor}
\usepackage[colorlinks=false,urlcolor=blue]{hyperref}
\usepackage{bbm}
\usepackage{subcaption}
\usepackage{multirow}
\usepackage{tabularx}
\usepackage{algorithm,algpseudocode}
\usepackage{tabulary}
\usepackage{setspace}

\usepackage{fancyhdr}
\hypersetup{pdfstartview={XYZ null null 1.00}}

\title{Dynamic Question Ordering in Online Surveys\\\small (In submission to Journal of Official Statistics)}
\author{Kirstin Early\\Machine Learning Department\\Carnegie Mellon University\\5000 Forbes Avenue\\Pittsburgh, PA, 15213\\\texttt{kearly@cs.cmu.edu}
\and
Jennifer Mankoff\\Human-Computer Interaction Institute\\Carnegie Mellon University\\5000 Forbes Avenue\\Pittsburgh, PA, 15213\\\texttt{jmankoff@cs.cmu.edu}
\and
Stephen Fienberg\\Department of Statistics\\Carnegie Mellon University\\5000 Forbes Avenue\\Pittsburgh, PA, 15213\\\texttt{fienberg@stat.cmu.edu}
}
\date{}

\newcommand \R {\mathbb{R}}
\newcommand \Zp {\mathbb{Z}^+}
\newcommand \E [1]{\mathbb{E}\lbrack #1 \rbrack}

\newcommand{\argmin}[2][]{
  \ensuremath{\arg\underset{#2}{\min}\, #1}}
\newcommand{\eg}{\emph{e.g.}, }
\newcommand{\ie}{\emph{i.e.}, }

\makeatletter
\let\OldStatex\Statex
\renewcommand{\Statex}[1][3]{%
  \setlength\@tempdima{\algorithmicindent}%
  \OldStatex\hskip\dimexpr#1\@tempdima\relax}
\makeatother

\makeatletter
\newlength{\continueindent}
\renewenvironment{algorithmic}[1][0]%
   {%
   \edef\ALG@numberfreq{#1}%
   \def\@currentlabel{\theALG@line}%
   \setcounter{ALG@line}{0}%
   \setcounter{ALG@rem}{0}%
   \let\\\algbreak%
   \expandafter\edef\csname ALG@currentblock@\theALG@nested\endcsname{0}%
   \expandafter\let\csname ALG@currentlifetime@\theALG@nested\endcsname\relax%
   \begin{list}%
      {\ALG@step}%
      {%
      \rightmargin\z@%
      \itemsep\z@ \itemindent\z@ \listparindent2em%
      \partopsep\z@ \parskip\z@ \parsep\z@%
      \labelsep 0.5em \topsep 0.2em
      \ifthenelse{\equal{#1}{0}}%
         {\labelwidth 0.5em}%
         {\labelwidth 1.2em}%
       \leftmargin\labelwidth \addtolength{\leftmargin}{\labelsep}
      \ALG@tlm\z@%
      }%
      \parshape 2 \leftmargin \linewidth \continueindent \dimexpr\linewidth-\continueindent\relax
   \setcounter{ALG@nested}{0}%
   \ALG@beginalgorithmic%
   }%
   {
   \ALG@closeloops%
   \expandafter\ifnum\csname ALG@currentblock@\theALG@nested\endcsname=0\relax%
   \else%
      \PackageError{algorithmicx}{Some blocks are not closed!!!}{}%
   \fi%
   \ALG@endalgorithmic%
   \end{list}%
   }%
\makeatother
\algnewcommand{\LineComment}[1]{\State \(\triangleright\) #1}

\begin{document}
\maketitle

\doublespacing
\begin{abstract}
\begin{quote}
Online surveys have the potential to support adaptive questions, where later questions depend on earlier responses. Past work has taken a rule-based approach, uniformly across all respondents. We envision a richer interpretation of adaptive questions, which we call dynamic question ordering (DQO), where question order is personalized. Such an approach could increase engagement, and therefore response rate, as well as imputation quality. We present a DQO framework to improve survey completion and imputation. In the general survey-taking setting, we want to maximize survey completion, and so we focus on ordering questions to engage the respondent and collect hopefully all information, or at least the information that most characterizes the respondent, for accurate imputations. In another scenario, our goal is to provide a personalized prediction. Since it is possible to give reasonable predictions with only a subset of questions, we are not concerned with motivating users to answer all questions. Instead, we want to order questions to get information that reduces prediction uncertainty, while not being too burdensome. We illustrate this framework with an example of providing energy estimates to prospective tenants. We also discuss DQO for national surveys and consider connections between our statistics-based question-ordering approach and cognitive survey methodology.
\end{quote}
\end{abstract}

\section{Introduction}
\label{sec:intro}
Survey response rates have been falling for decades, leading to results that do not necessarily represent the full population of interest~\cite{porter2004raising}. Online surveys tend to have much lower response rates than traditional mail-out/mail-back and telephone surveys~\cite{shih2008comparing}.  Unlike  these traditional-styled surveys, online surveys can easily support adaptive question ordering, where the order of later questions depends on responses to earlier questions. Past work in adaptive questions for online surveys has taken a rule-based, question-specific approach where a certain response to a certain question leads to a new set of questions, uniformly across all respondents (\eg~\cite{pitkow1995using,bouamrane2008gathering}). We envision a richer interpretation of adaptive question ordering, where question order is dynamic and personalized to the individual respondent, depending on their previous answers. Such a dynamic question-ordering approach has the potential of increasing engagement, and therefore response rates, as well as the quality of imputation for missing values.

In addition to the use of surveys to gain insights about general populations, we can use survey results to give useful information to individual respondents.  We consider a respondent's answering a sequence of questions to receive a personalized estimate as a type of survey too. An example of such a survey is a carbon calculator, where a user provides information about their home infrastructure and energy consumption to get an estimate of their carbon footprint~\cite{pandey2011carbon}. Here the respondent receives useful information from the survey, has a personal incentive to complete the survey, and likely self-selects into the survey.   It is possible that the user does not need to answer all questions to get an accurate estimate of their carbon footprint; for some users, certain features will be more relevant than for other users. We can lower the cost (to users) of providing answers by ordering the questions so that the most informative questions for a particular user are asked first.

We present a general framework for dynamically ordering the questions that make up a survey questionnaire, based on previous responses, to engage respondents, and improve survey completion and imputation of unknown items. Our work considers two scenarios for data collection from survey-takers. In the first, we want to maximize survey completion (and the quality of necessary imputations) and so we focus on ordering questions to engage the respondent and collect hopefully all the information we seek, or at least the information that most characterizes the respondent so imputed values will be accurate. In the second scenario, our goal is to give the respondent a personalized prediction, based on information they provide. Since it is possible to give a reasonable prediction with only a subset of questions, we are not concerned with motivating the user to answer all questions. Instead, we want to order questions so that the user provides information that most reduces the uncertainty of our prediction, while not being too burdensome to answer.

Any statistics-based approach to dynamic question ordering of the sort we consider here would seem to run counter to traditional arguments that questionnaires should have a fixed structure for all respondents and when the same quantities, \eg unemployment or poverty, are measured by surveys over time. Just over thirty years ago, the cognitive aspects of survey methodology (CASM) movement, \eg see \cite{jabine1984CASM,sudman1996thinking,tanur1992questions},  made the argument that this traditional approach to survey design shackled respondents and often prevented them from providing the very answers  that the survey methodologists sought for their questions, \eg see \cite{suchman1990interactional, tanur1992questions}. We believe our approach reopens the door to the arguments raised by that movement, but in a very different manner, and somehow survey statisticians will ultimately need to blend the lessons from the CASM movement with the needs for cost-driven dynamic ordering.

The remaining sections of the paper are as follows:  first, in Section~\ref{sec:rel-work}, we  review related work in question ordering from a variety of fields. Then, in Section~\ref{sec:recs-dqo}, we jump into a specific formulation of dynamic question ordering (DQO) from a project to provide personalized energy estimates to prospective tenants, drawing on data from the Residential Energy Consumption Survey~\cite{recs2009}. Our approach makes useful, individualized predictions of energy usage at under 30\% of the cost of the full-feature model.  In Section~\ref{sec:gen-framework} we formalize and generalize the DQO framework, beyond the particular application in the previous section. In Section~\ref{sec:other-apps} we set forth a broader view of the forms dynamic question ordering can take in other national surveys and suggest how they might benefit from a dynamic question-ordering approach. Finally, in Sections~\ref{sec:future} and~\ref{sec:concl} we summarize our contribution and note avenues for future work in this area.

\section{Some Related Work}
\label{sec:rel-work}
There is a rich literature focusing on  adaptively ordering questions to improve outcomes while minimizing respondent burden, across multiple fields. Examples include adaptive design in survey methodology, adaptive treatment design in medical statistics, adaptive testing in educational research, and test-time feature selection in machine learning.

\subsection{Adaptive Survey Design}
Adaptive survey design (ASD) attempts to improve survey quality (in terms of achieving a higher response rate or lower error) by giving respondents custom survey designs, rather than the same one~\cite{schouten2013optimizing}. Usually ASD tries to minimize nonresponse, and designs involve factors like number of follow-ups, which can be costly. The general technique is to maximize survey quality while keeping costs below a budget. 

Often in ASD, changes in survey design happen between \emph{phases} of the survey, where the exact same survey protocol (\eg sampling frame, survey mode, measurement conditions) is in place within a phase and results from that phase inform changes to the protocol for the next phase. Groves and Heeringa~\cite{groves2006responsive} introduce an approach they call responsive survey design, which uses indicators of the cost and error of design features to make decisions about how to change the survey design in future phases and then combines data from all phases into a final estimator. They also introduce the concept of \emph{phase capacity}\textemdash once a stable estimate has been reached in a design phase, it is unlikely that expending more effort in that phase will result in a better estimate. Their definition of ``effort" focuses on collecting participants for each phase. They propose the use of error-sensitive indicators to identify when a phase has reached capacity and no more participants need to be recruited for that phase. This notion of phase capacity could extend to reaching a stable estimate of a participant's survey-answering, and no more questions need to be asked.

\subsection{Adaptive Treatment Strategies}
In the field of medical statistics, adaptive treatment strategies (also called dynamic treatment regimes) continually adjust treatments, according to decision rules, depending on an individual's responses to previous treatments as well as characteristics of the patient~\cite{collins2007multiphase}. This technique contrasts with the research standard of randomized controlled trials, but more closely matches real-world practice of medical intervention (since, when a treatment fails for a particular patient, that patient is reassigned to a new treatment, based on how they reacted). Adaptive treatment strategies are targeted for an \emph{individual}, rather than basing future treatment decisions on outcomes of previous patients.

The design of the sequential multiple assignment randomized (SMAR) trial~\cite{murphy2005experimental} chooses a decision to make at each point according to what action will maximize the expected treatment outcome, given past information that has occurred. SMAR trials randomize individuals to different treatments at each decision time point.

Adaptive treatment strategies have been applied to treat depression, with the STAR*D (sequenced treatment alternatives to relieve depression) treatment~\cite{rush2004sequenced}, in which patients who did not respond to less-intensive therapies were randomly assigned to more intensive treatments at higher levels; to treat schizophrenia, with the CATIE (clinical antipsychotic trials of intervention effectiveness) design~\cite{stroup2003national}, a three-phase study where patients were randomly assigned to new treatments at successive phases if they did not respond to earlier treatments; to treat advanced prostate cancer~\cite{wang2012evaluation} by randomizing nonfavorably-responding patients to untried chemotherapy treatments at eight-week intervals, up to four times; and many other medical settings (\eg smoking cessation~\cite{collins2005strategy}, pediatric generalized anxiety disorders~\cite{almirall2012designing}, and mood disorders~\cite{lavori2000flexible,kilbourne2014protocol}).


\subsection{Adaptive Testing}
For tests that measure ability or aptitude, adaptive testing selects test questions based on the respondent's answers to previous questions. The goal is to measure the examinee's achievement accurately, without making the examinee answer too many questions. Adaptive tests have been shown to be as reliable and valid as conventional tests (with static question orders), while reducing test length up to 50\%~\cite{weiss1982improving}. Unlike classical test theory, which assumes all questions equally indicate an assessment outcome, item response theory (IRT) ~\cite{lord1980applications} considers \emph{individual} test questions through an item response function, the probability of a correct answer by an individual at a particular skill level $\theta$. The item response function has three parameters: the pseudo-chance score level (how easy it is to guess the correct answer), item difficulty (how hard it is to answer the question), and discriminating power (how much the skill level influences question response). According to Weiss~\cite{weiss1982improving}, an IRT-based adaptive testing framework has the following three components: (1) a way to choose the first item to ask, (2) a way to score items and choose the next item to ask during test administration, and (3) a way to choose to end the test, based on an individual's performance.

Weiss and Kingsbury~\cite{weiss1984application} introduce adaptive mastery testing to assess a student's achievement level $\hat{\theta}$, specifically how the estimated achievement level compares to a ``mastery level," $\theta_m$. At each time point, a question is selected which gives the maximum information at the student's current estimated mastery level and asked. As the student answers questions, the estimate $\hat{\theta}$ is updated, along with a confidence interval. Once the confidence interval for $\hat{\theta}$ no longer includes $\theta_m$, the test is finished and the student's mastery level is assigned as sufficient or not (depending if $\theta_m$ lies above or below the confidence interval for $\hat{\theta}$).

More recently, IRT-based adaptive testing has been used for diagnoses of mental health disorders through patient questionnaires~\cite{gibbons2016computerized}. Their experiments show that their adaptive diagnosis process can, in only one minute of testing, arrive at the same diagnosis as a trained clinician in one hour. Montgomery and Cutler~\cite{montgomery2013computerized} have also used IRT-based adaptive testing, but for public opinion surveys. In an empirical study using adaptive testing to measure respondents' political knowledge, the authors found that the adaptive testing approach could produce more accurate measurements than traditional test administration, at a 40\% reduction in questionnaire length.


\subsection{Test-time Feature Selection}

In the case where survey collection is targeted toward the goal of providing the user with a personalized prediction (\eg for energy consumption), at test time, the goal is to make a prediction on a new example. Making a prediction on a test instance requires gathering feature values, which can be costly, especially if it requires cooperation from users who might stop before completion. In this case, strategically ordering questions asked (based on previous answers) can get the most useful information first, while providing predictions on partial information. This way, people receive meaningful predictions without spending much time or effort answering questions.

The test-time feature ordering problem resembles active learning, which assumes \emph{labels} are expensive. Active learning algorithms strategically select which unlabeled points to query to maximize the model's performance (using both labeled and unlabeled data) while minimizing the cost of data collection~\cite{cohn1996active}. Test-time feature ordering has a similar goal of making accurate predictions while keeping data collection costs low; however, rather than choosing an \emph{example to be labeled}, test-time feature ordering chooses a \emph{single feature to be entered} (by asking the user a question). 

He, Daum\'{e} III, and Eisner~\cite{he2012cost} consider the setting of test-time feature selection, where all features are available for training, and at test time they want an instance-specific subset of features for prediction, trading off feature cost with prediction accuracy. They formulate dynamic feature selection as a Markov decision process (MDP). The policy selects a feature to add; the reward function reflects the classifier margin with the next feature, penalized by the cost of including that feature. However, this method does not make sequential predictions, and instead only chooses whether to keep getting features or to stop and make a final prediction. Karayev, Baumgartner, Fritz, and Darrell~\cite{karayev2012timely} also take an MDP approach to classify images with a framework they call ``timely object recognition," which sequentially runs detectors and classifiers on subsets of the image and uses previous results to inform the next action to take, while providing the best object recognition in the available runtime.


Most work in test-time feature ordering does not consider the situation of providing predictions with partial information \emph{as} questions are answered, nor does it address the issue of giving measures of prediction uncertainty to users.



\section{Providing Personalized Energy Estimates with the Residential Energy Consumption Survey}
\label{sec:recs-dqo}
In this section, we illustrate the concept of dynamic question ordering for prediction with an application of predicting energy consumption for a prospective tenant in a potential home.

Selecting homes with energy-efficient infrastructure is important for renters, because infrastructure influences energy consumption far more than in-home behavior~\cite{dietz2009household}. The importance of energy estimates for apartment hunters is twofold. First, since renters often cannot make infrastructure upgrades for efficiency in a property they do not own, they need to know upfront the expected costs of living in a rental unit. Second, 30\% of the U.S. population rent, and renters move on average every two years~\cite{ahs2013}; therefore, renters can potentially choose improved infrastructure more frequently than we can expect homeowners to make costly upgrades.

Personalized energy estimates can guide prospective tenants toward energy-efficient homes, but this information is not readily available. Utility estimates are not typically offered to house-hunters, and existing technologies like carbon calculators require users to answer (prohibitively) many questions that may require considerable research to answer. For the task of providing personalized utility estimates to prospective tenants, we present a cost-based model for feature selection at training time, where all features are available and costs assigned to each feature reflect the difficulty of acquisition. At test time, we have immediate access to some features but others are difficult to acquire (costly). In this limited-information setting, we strategically order questions we ask each user, tailored to previous information provided, to give the most accurate predictions while minimizing the cost to users. During the critical first 10 questions that our approach selects, prediction accuracy improves equally to fixed-order approaches, but prediction certainty is higher~\cite{early2016dynamic}.

\subsection{Introduction}
Since energy consumption depends on home infrastructure (\emph{e.g.}, square footage) and occupant behavior (\emph{e.g.}, preferred temperature), we can learn the relationship between these features and energy consumption through established datasets, like the Residential Energy Consumption Survey (RECS)~\cite{recs2009}. Some information can be extracted automatically from online rental advertisements, while other information must be provided by prospective tenants at various costs (for example, the question of how many windows a home has requires more effort to answer than how many people will live there). 
To develop a predictive model of energy consumption at training time, we begin with the extractable (\emph{i.e.}, ``free") features in a regression model to predict energy usage and use forward selection to add a subset of the costly features.

After learning this predictive model on the training dataset, the main problem lies in how to make a prediction on a new test instance. Initially, only a subset of the features (the free features) in the model are available, and asking users for each unknown value incurs a cost, depending on how hard it is to provide that feature. Our dynamic question-ordering algorithm (DQO) chooses the best question to ask next by considering which feature, if its value were known, would most reduce uncertainty, measured by the width of the prediction interval, with a penalty term on that feature's cost.

\subsection{Method}
\subsubsection{Training Time: Cost-Aware Feature Selection}
A greedy approximation to feature selection, forward selection starts with an empty feature set and, at each iteration, adds the feature that minimizes error~\cite{harrell2001regression,tropp2004greed}.
For this analysis, we started with the free extractable features (rather than no features, as in classic forward selection) and minimized leave-one-out cross-validation error with linear regression to add successive higher-cost features. 

\subsubsection{Test Time: Cost-Effective Dynamic Question Ordering}
After learning regression models for energy usage on the selected features from our training set, we want to make a prediction on a new test point, where initially only some features of the model are available. Our approach considers a trajectory of prediction intervals as a user provides information. A prediction interval consists of a lower and upper bound such that the true value lies in this interval with at least some probability~\cite{weisberg2014applied}. Prediction interval width corresponds to prediction uncertainty: a wider interval means less confidence. We select as the optimal next question the one whose inclusion most reduces the expected value of the prediction interval width; that is, it most reduces the expected uncertainty of the next prediction.

In this problem, there are features that are unknown (not yet supplied by the user). We use $k$ nearest neighbors ($k$NN) \cite{cover1967nearest} to supply values for unanswered features in vector $x\in\R ^d$. For each unknown feature $f$, we find the $k$ data points in the training set $X\in\R^{n\times d}$ ($n$ samples, each $d$-dimensional) that are closest to $x$, along the dimensions $\mathcal{K}$ 
that are currently known. Then we estimate $x_f$ as $z_f$, the mean or mode, as appropriate, of feature $f$ in the $k$ nearest neighbors (depending whether the feature is continuous or discrete) 
(see Algorithm~\ref{alg:knn-update}).
\begin{algorithm}
\caption{Estimating values $z$ for still-unknown features}
\label{alg:knn-update}
\begin{algorithmic}[1]
\Require $X\in\R^{n\times d},x\in\R^d,\mathcal{K}\subseteq \lbrace 1,...,d\rbrace,k\in\Zp$
\Ensure $z\in\R^d$
\Function{estimate\_features}{$X,x,\mathcal{K},k$}
\State $z_{\mathcal{K}}\gets x_{\mathcal{K}}$ \Comment{Copy over the known features}
\State $\mathcal{I} \gets \text{get\_knn}(X_{:,\mathcal{K}},z_\mathcal{K},k)$ \Comment{Index $z_\mathcal{K}$'s $k$NNs}
\For {$f\in\lbrace 1,...,d\rbrace \setminus\mathcal{K}$} \Comment{For all unknown $f$}
\State $z_f \gets \text{mean}(X_{\mathcal{I},f})$ \Comment{Estimate $z_f$ from $k$NNs' values for feature $f$}
\EndFor
\State \textbf{return} $z$
\EndFunction
\end{algorithmic}
\end{algorithm}

Because these $z$ values estimate unknown features of $x$, we use the measurement error model (MEM)~\cite{fuller2009measurement} to capture error associated with estimated features. Unlike traditional regression models, MEMs do not assume we observe each component $x_f$ exactly; there is an error $\delta_f$ associated with the estimation: 
\begin{equation*}
\label{eqn:mem-obs}
z_f = x_f + \delta_f, \;\text{where}\; \E{\delta_f|x_f} = 0. 
\end{equation*}

Prediction $\hat{y}$ still depends on the \emph{true, unobserved} value $x$: 
\begin{equation*}
\hat{y} = \hat{\beta}^T\bar{x} = \hat{\beta}(\bar{z} - \bar{\delta}),
\end{equation*}
where $\hat{\beta}\in\R ^{d+1}$ is the parameter vector learned on the training set $X$ (recall all feature values are known at training time). The notation $\bar{x},\bar{z},\bar{\delta}$ means vectors $x,z$ have a 1 appended to them and $\delta$ a 0 to account for the constant term in the regression. Let $\bar{X}$ extend this notion to the training matrix: $\bar{X} = \lbrack \mathbf{1}^n X\rbrack$. 

We can calculate a $100(1-\alpha)\%$ prediction interval for a new point $z$ as
\begin{equation}
\begin{aligned}
\label{eqn:mem-lr-pred-int}
\hat{y} \pm t_{n-d-1;\alpha/2}
\sqrt{\hat{\sigma}^2\left(1+\bar{z}^T(\bar{X}^T\bar{X})^{-1}\bar{z} + \bar{\delta}^T(\bar{X}^T\bar{X})^{-1}\bar{\delta}\right)},
\end{aligned}
\end{equation}
where the $\bar{\delta}^T(\bar{X}^T\bar{X})^{-1}\bar{\delta}$ term accounts for error from estimated features and $t_{n-d-1;\alpha/2}$ is the value at which a Student's $t$ distribution with $n-d-1$ degrees of freedom has cumulative distribution function value $\alpha/2$. 
We can estimate $\delta$ from training data by calculating the error of predicting each feature with $k$NN, from the other features. We also estimate $\hat{\sigma}^2$, the regression variance, from training data.

Then, we cycle through each candidate feature $f$ and compute the expected prediction interval width $\E{w(f)}$ for asking that feature next, over each value $r$ that feature $f$ might take on from its range of potential values $R$:

\begin{equation}
\label{eqn:e-pred-width}
\begin{aligned}
\E {w(f)} = 2\cdot t_{n-d-1;\alpha/2}\sum\limits_{r\in R} p(z_f = r)
  \sqrt{
    \begin{aligned}
     \hat{\sigma}^2\biggl( 1+ \bar{z}_{f:=r}^T (\bar{X}^T\bar{X})^{-1}\bar{z}_{f:=r}\; + 
      \bar{\delta}_{f:=0}^T(\bar{X}^T\bar{X})^{-1}\bar{\delta}_{f:=0}\biggr)
    \end{aligned}
  },
\end{aligned}
\end{equation}
where $p(z_f = r)$, the probability that the $f$-th feature's value is $r$, is calculated empirically from the training set, and the notation $\bar{u}_{f:=q}$ means the $f$-th component of $u$ is replaced with the value $q$. Algorithm~\ref{alg:e-pred-int} writes this process in pseudocode. In this algorithm, feat\_ranges and feat\_proportions are both $d$-dimensional cell arrays where the $f$-th cells contain, respectively, the set of values $R$ that feature $f$ can take on and the proportions $p\in\R^{|R|}$ that each value $r\in R$ appears in the training set. The output, $E\in\R^d$, is a vector where the $f$-th component is the expected prediction interval width if the value of feature $f$ were known.

\begin{algorithm}[H]
\caption{Calculating the expected prediction interval width for each candidate feature to be asked}
\label{alg:e-pred-int}
\begin{algorithmic}[1]
\Require $X\in\R^{n\times d}$, $\mathcal{K}\subseteq \lbrace 1,...,d\rbrace$, $z\in\R^{d}$, $\delta\in\R^{d}$, $\hat{\sigma}^2\in\R$, $\alpha\in\lbrack 0,1\rbrack$, $\text{feat\_ranges}$, $\text{feat\_proportions}$
\Ensure $E\in\R^{d}$
\Function{Expected\_interval\_width}{$X$, $\mathcal{K}$, $z$, $\delta$, $\hat{\sigma}^2$, $\alpha$, $\text{feat\_ranges}$, $\text{feat\_proportions}$}
\State $v\gets\mathbf{0}^d$
\State $\bar{X}\gets \lbrack \mathbf{1}^n X\rbrack$
\For {$f\in\lbrace 1,...,d\rbrace \setminus\mathcal{K}$} \Comment{For all unknown $f$}
\State $R\gets \text{feat\_ranges}\lbrace f\rbrace,\, u\gets \mathbf{0}^{|R|}$
\For {$\ell \in\lbrace 1,...,|R|\rbrace$} \Comment{For each value $f$ can take on}
\State $\bar{z}\gets \lbrack 1; z\rbrack,\, \bar{z}_{f+1}\gets R_\ell$ \Comment{$f$-th feature is assigned}
\State $\bar{\delta}\gets \lbrack 0; \delta\rbrack,\, \bar{\delta}_{f+1}\gets 0$ \Comment{No uncertainty in $f$-th feature}
\State $u_\ell \gets \bar{z}^T(\bar{X}^T\bar{X})^{-1}\bar{z} + \bar{\delta}^T(\bar{X}^T\bar{X})^{-1}\bar{\delta}$
\EndFor
\State $p\gets\text{feat\_proportions}\lbrace f\rbrace$
\State $v_f\gets p^Tu$
\EndFor
\State \textbf{return} $E\gets 2\cdot t_{n-d-1;\alpha/2}\sqrt{\hat{\sigma}^2(1+v)}$ \Comment{Elementwise operation on $v$}
\EndFunction
\end{algorithmic}
\end{algorithm}

\begin{algorithm}[H]
\caption{Dynamically choosing a question ordering $\mathcal{A}$ and making a sequence of predictions $\hat{y}$ at the current feature values and estimates as feature values are provided}
\label{alg:dqo-all}
\begin{algorithmic}[1]
\Require $X\in\R^{n\times d}$, $x\in\R^{d}$, $\mathcal{K}\subseteq \lbrace 1,...,d\rbrace$, $k\in\Zp$, $\delta\in\R^{d}$, $\alpha\in\lbrack 0,1\rbrack$, $\text{feat\_ranges}$, $\text{feat\_proportions}$, $\hat{\beta}\in\R^{d+1}$, $\hat{\sigma}^2\in\R$, $\lambda\in\R$, $c\in\R^{d}$
\Ensure $\mathcal{A}\subseteq\lbrace 1,...,d\rbrace, \hat{y}\in\R^{|\mathcal{A}|+1}$
\Function{DQO\_all}{$X$, $x$, $\mathcal{K}$, $k$, $\delta$, $\alpha$, $\text{feat\_ranges}$, $\text{feat\_proportions}$, $\hat{\beta}$, $\hat{\sigma}^2$, $\lambda$, $c$}
\State $\mathcal{A}\gets \lbrace\rbrace, \, \hat{y}\gets \lbrace\rbrace$
\For {$i\in\lbrace 1,...,d-|\mathcal{K}|\rbrace$}
\State $z\gets \Call{estimate\_features}{X,x,\mathcal{K},k}$
\State $\hat{y}_i\gets \hat{\beta}^Tz$ \Comment{Predict on features and estimates}
\State $E\gets$ \textproc{Expected\_interval\_width}(
\Statex[5] $X$, $\mathcal{K}$, $z$, $\delta$, $\hat{\sigma}^2$, $\alpha$, $\text{feat\_ranges}$, $\text{feat\_proportions}$)
\State $f^\star\gets \argmin{f\notin\mathcal{K}}{\left(E_f + \lambda\cdot c_f\right)}$
\State $\mathcal{A}\gets\mathcal{A}\cup\lbrace f^\star\rbrace, \mathcal{K}\gets\mathcal{K}\cup\lbrace f^\star\rbrace$
\State $z_{f^\star}\gets x_{f^\star}$ \Comment{Ask and receive value for $f^\star$}
\State $\delta_{f^\star}\gets 0$ \Comment{No more uncertainty in $f^\star$}
\EndFor
\State $z\gets \Call{estimate\_features}{X,x,\mathcal{K},k}$
\State $\hat{y}_{d-|\mathcal{K}|+1}\gets\hat{\beta}^Tz$ \Comment{Make final prediction}
\State \textbf{return} $\mathcal{A},\hat{y}$
\EndFunction
\end{algorithmic}
\end{algorithm}

Including the feature that attains the narrowest expected prediction interval width $\E {w(f)}$ will reduce the uncertainty of our prediction more than any other feature. This approach allows incorporation of feature cost into the question selection, by weighting the expected prediction interval width against the cost of acquiring the feature:
\begin{equation*}
f^\star = \argmin{f}{\left(\E{w(f)} + \lambda\cdot c_f\right)},
\end{equation*}
where $c_f$ is the cost of feature $f$ and $\lambda\in\R$ trades off feature cost with reduced uncertainty. A high-cost feature might not be chosen, 
if another feature can provide enough improvement at lower cost. We ask for this information, update our vector of known data with the response (and estimate the unknown features again, now including the new feature in the set for $k$NN prediction), and repeat the process until all feature values are filled in (or the user stops answering). 
Algorithm~\ref{alg:dqo-all} formalizes this dynamic question-ordering (DQO) process.

More generally, this algorithm can be seen as a framework that makes predictions on partial information and selects which feature to query next by (1) estimating values for unknown features (here with $k$NN) and (2) asking for the feature that will most reduce the expected uncertainty of the next prediction (here measured by prediction interval width). With this approach, we strategically order questions, tailored to previous information, to give accurate predictions while minimizing the user burden of answering many or difficult questions that will not provide a substantial reduction in prediction uncertainty.

\subsection{Data}
The Residential Energy Consumption Survey (RECS) contains information about home infrastructure, occupants, and energy consumption. We can use this dataset to learn relationships between household features and energy consumption to predict energy usage for prospective tenants. The most recently released RECS was a nationally representative sample of 12,083 homes across the U.S.~\cite{recs2009}. For each household, RECS records fuel consumption by fuel type (\emph{e.g.}, electricity, natural gas) and around 500 features of the home (\emph{e.g.}, age of refrigerator, number of occupants).

\subsubsection{Defining Feature Costs}
In our problem setting, features have different costs of obtaining, and we want to build models and make predictions that leverage features cost-effectively. Some information is easily found in the rental listing (\emph{e.g.}, number of bedrooms) and is therefore ``free," while other information requires asking users. For example, the number of windows does not appear in listings and would require a prospective tenant to visit each site;  consequently, this question has high cost. Other useful features relate to occupant behavior (\emph{e.g.}, preferred temperature). These questions likely remain constant for each user across homes and therefore require asking only once and are cheaper. We categorize feature costs as ``extractable/free" (can be automatically extracted from rental listings), ``low" (occupant-related; require asking only once), and ``high" (unit-related; must be answered once for each apartment and may require a site visit). 
Table~\ref{tab:extr-feats} lists the information used for extractable features and how often it appears in Rent Jungle, a company that scrapes rental listings from the internet; these ``free" features appear in the majority of listings on Rent Jungle.

\begin{table}[h]
\caption{The features we define as extractable (\emph{i.e.}, ``free") appear in most of the listings on Rent Jungle. Geographic features associated with the city, zip code, or state include climate zone and whether the area is urban or rural, among others.}
\label{tab:extr-feats}
\begin{center}
\begin{tabular}{|l|c|}
\hline
\bf Feature &\bf Presence in Rent Jungle database\\
\hline
Number of bedrooms & 85\%\\
Number of full bathrooms & 57\%\\
Studio apartment & 85\%\\
City or zip code & 99\%\\
State & 100\%\\
\hline
\end{tabular}
\end{center}
\end{table}

\subsection{Experimental Validation}
We validated our test-time feature ordering approach on the RECS dataset for predicting household electricity and natural gas consumption. We restricted our analysis to homes in the same climate zone as our planned deployment location in Pittsburgh, a subset of 2470 households in climate zone 2. We used 90\% of these homes for training and the remaining 10\% for testing. The training set was further subdivided into feature selection and cross-validation subsets. We trained separate models for predicting electricity consumption (on all homes in our climate zone) and for predicting natural gas consumption (on the 75\% of homes that use natural gas) with forward feature selection~\cite{harrell2001regression,tropp2004greed}. Due to the similarity of the results from electricity prediction and natural gas prediction, we show here results from only electricity prediction.

\subsubsection{Test Time: Cost-Effective Dynamic Question Ordering}
After learning regression models for electricity and natural gas prediction on the training set, we want to make predictions of energy usage for a new test point. We use our dynamic question-ordering framework (DQO) to make sequential predictions with partial, evolving information. To apply DQO, we first used the training set to choose parameters for imputing unknown features ($k$, from Algorithm~\ref{alg:knn-update}, \textproc{estimate\_features}) and to estimate the measurement error $\delta$ (from Algorithm~\ref{alg:e-pred-int}, \textproc{Expected\_interval\_width}). Then, we simulated question-asking on RECS to evaluate the performance of DQO for test-time feature ordering.

\paragraph{Parameter Selection and Estimation}
We used the training set to choose $k=100$ for imputing the values of features that have not yet been asked, based on the prediction performance of $k$NN for the higher-cost features. Then, we estimated the measurement error $\delta_f$ for each feature as the error from $k$NN on the training set.

\paragraph{Simulating the Question-Asking and -Answering Process with RECS}
We simulated the process of asking and answering questions on the testing subset of RECS by hiding the values for features that were not yet known. After we used DQO to choose a feature to acquire (line 7 in Algorithm~\ref{alg:dqo-all}, \textproc{DQO\_all}), we ``asked" this question and unveiled its value (line 9 in Algorithm~\ref{alg:dqo-all}) once it was ``answered."

\paragraph{Evaluating DQO Performance with Prediction Certainty, Error, and Cost}
We evaluated the performance of our cost-effective {\em DQO} algorithm in making sequential predictions with partial, evolving information on a held-out test set. For comparison, we implemented several baselines. 
The {\em Random} algorithm chooses a random question ordering for each sample, the {\em Fixed Decreasing} algorithm asks questions in decreasing order of feature measurement error $\delta_f$ (identical ordering for all samples), and the {\em Fixed Selection} algorithm asks questions in the order of forward selection in the training phase (also identical for all samples).
Finally, the {\em Oracle} chooses the next best feature according to the minimum true prediction interval width (calculated on the test sample using true feature values, rather than the \emph{expected} width as in Algorithm~\ref{alg:dqo-all}). We tested two versions of DQO and oracle: ordering additional features {\em without cost} and {\em with cost} (implemented as $\lambda = 0$ and $\lambda > 0$). 

We calculated several metrics related to the trajectory of prediction
performance and cost, for orderings given by the algorithms: 
\emph{DQO} and \emph{Oracle with} and \emph{without cost, Random, Fixed Decreasing}, and \emph{Fixed Selection}. 
We summarized prediction performance with the width of the current prediction interval (prediction \emph{certainty}) and the absolute value of the difference between the current prediction and the truth (prediction \emph{error}); we also measured the cumulative cost of all features asked at each step (prediction \emph{cost}). Table~\ref{tab:results}  summarizes the metric trajectories as areas under the curve\textemdash smaller values are better because they mean the algorithm spent less time in high uncertainty, error, and cost.

\begin{figure}[!htb]
\centering
\begin{subfigure}{0.49\textwidth}
  \includegraphics[width=\linewidth]{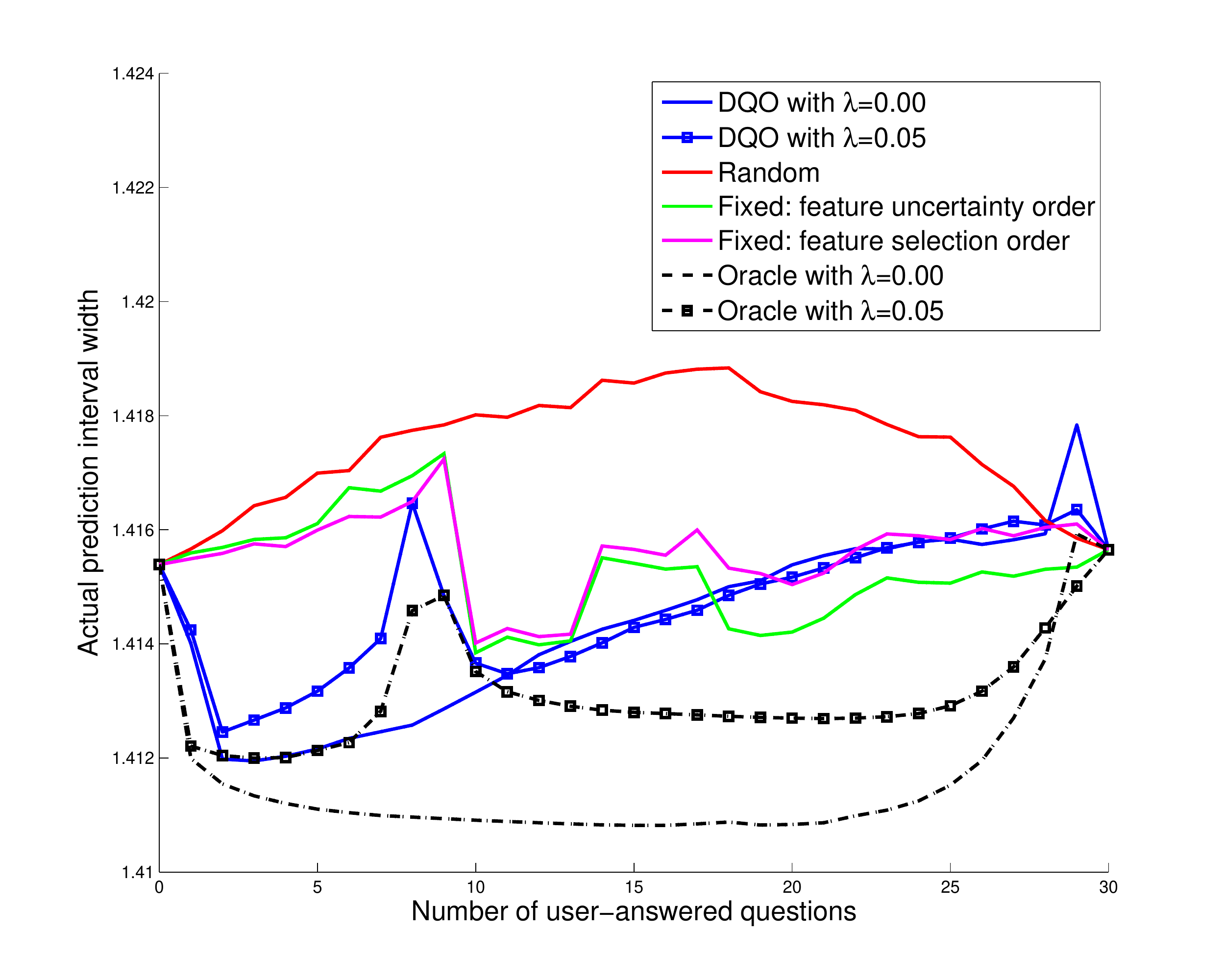}
  \caption{Prediction interval widths as questions are asked: DQO results in more certain predictions (\emph{i.e.}, lower prediction interval widths) than the baseline orderings.}
  \label{fig:pred-int-elec}
\end{subfigure}\hfill
\begin{subfigure}{0.49\textwidth}
  \includegraphics[width=\linewidth]{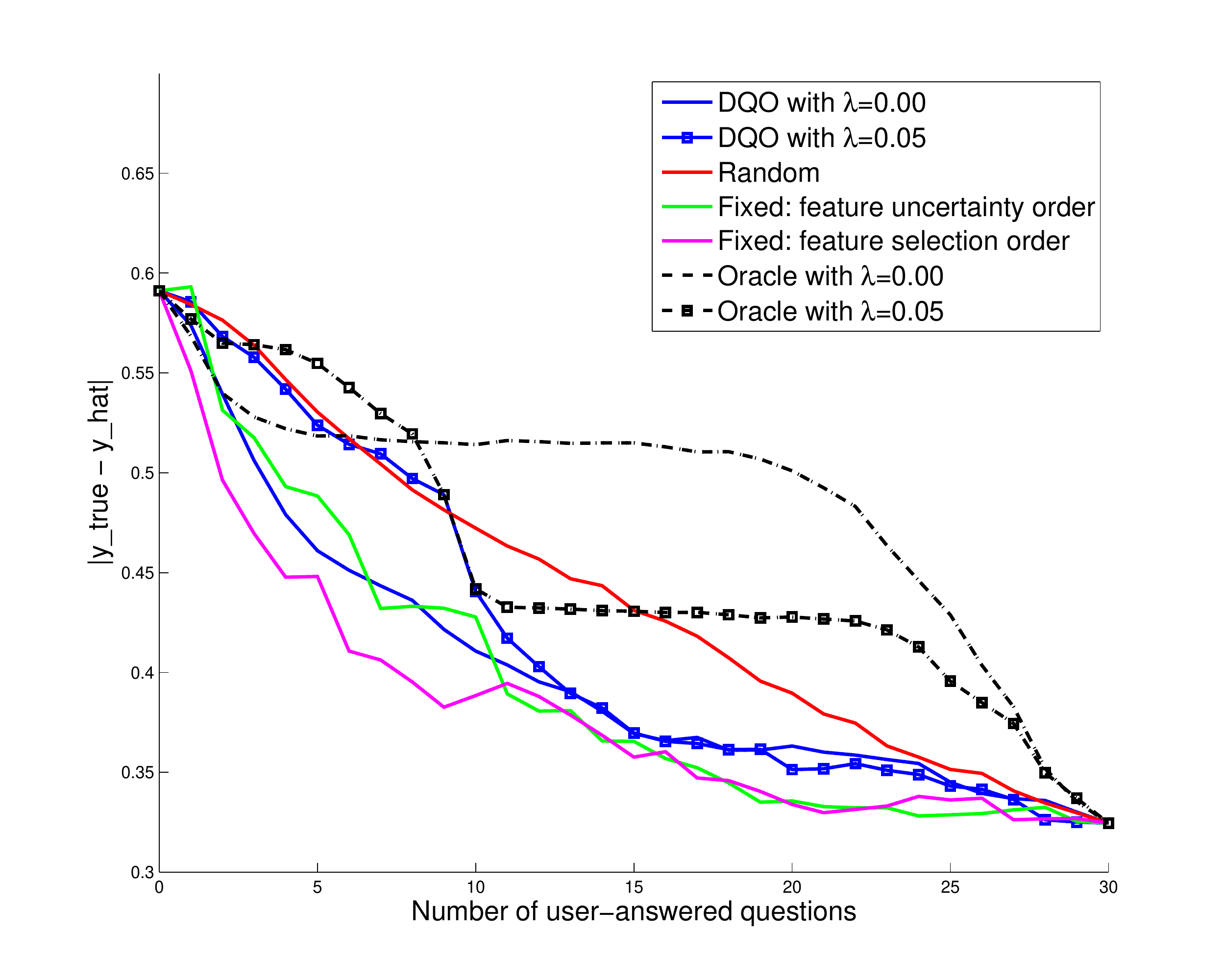}
  \caption{Mean absolute error as questions are asked: DQO results in similarly-correct predictions as baselines.}
  \label{fig:ydiff-elec}
\end{subfigure}\hfill
\begin{subfigure}{0.49\textwidth}
  \includegraphics[width=\linewidth]{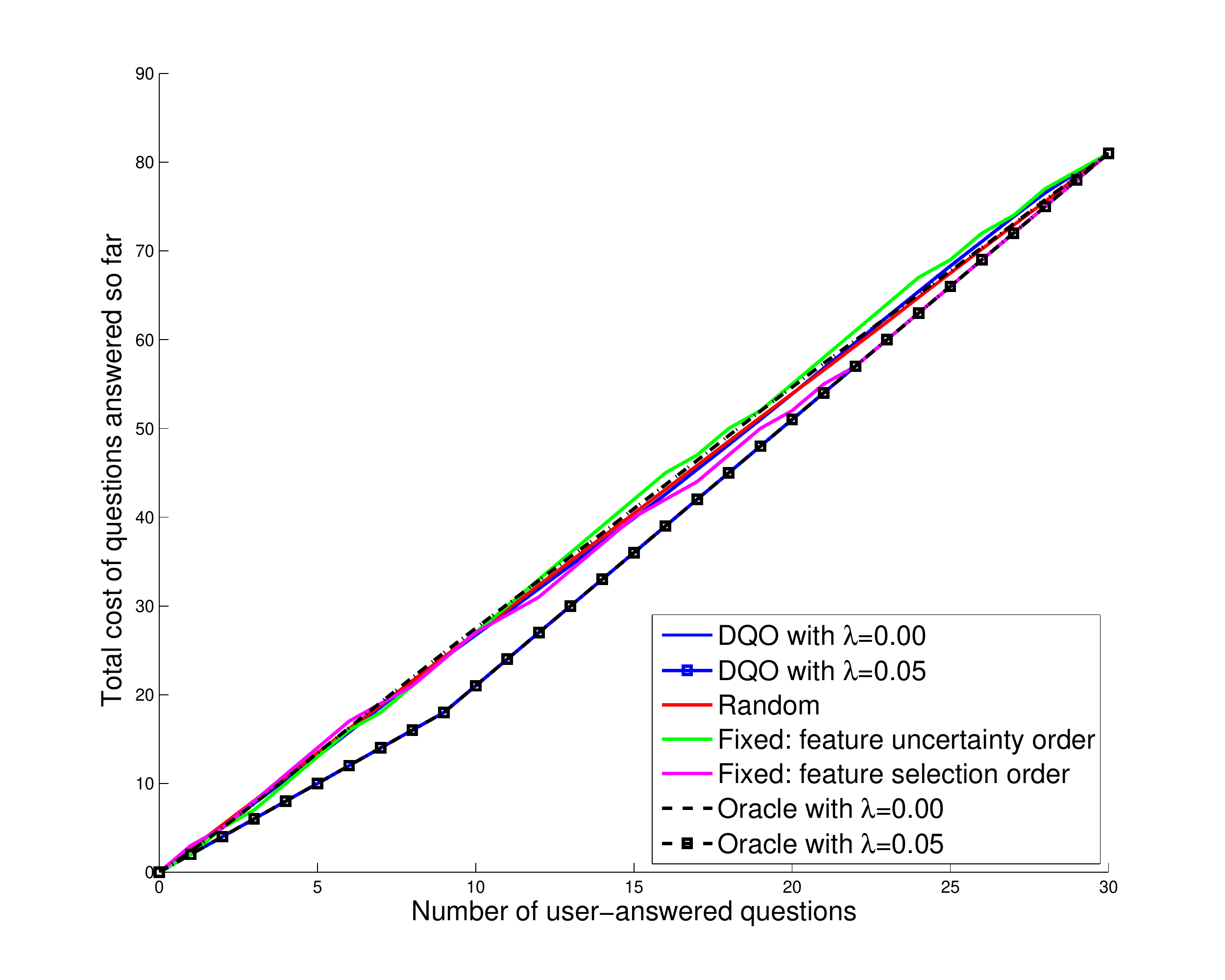}
\caption{Total feature costs as questions are asked: the tradeoff parameter $\lambda$ influences when expensive features are included.}
\label{fig:prog-cost-elec}
\end{subfigure}\hfill
\begin{subfigure}{0.49\textwidth}
 \includegraphics[width=\columnwidth]{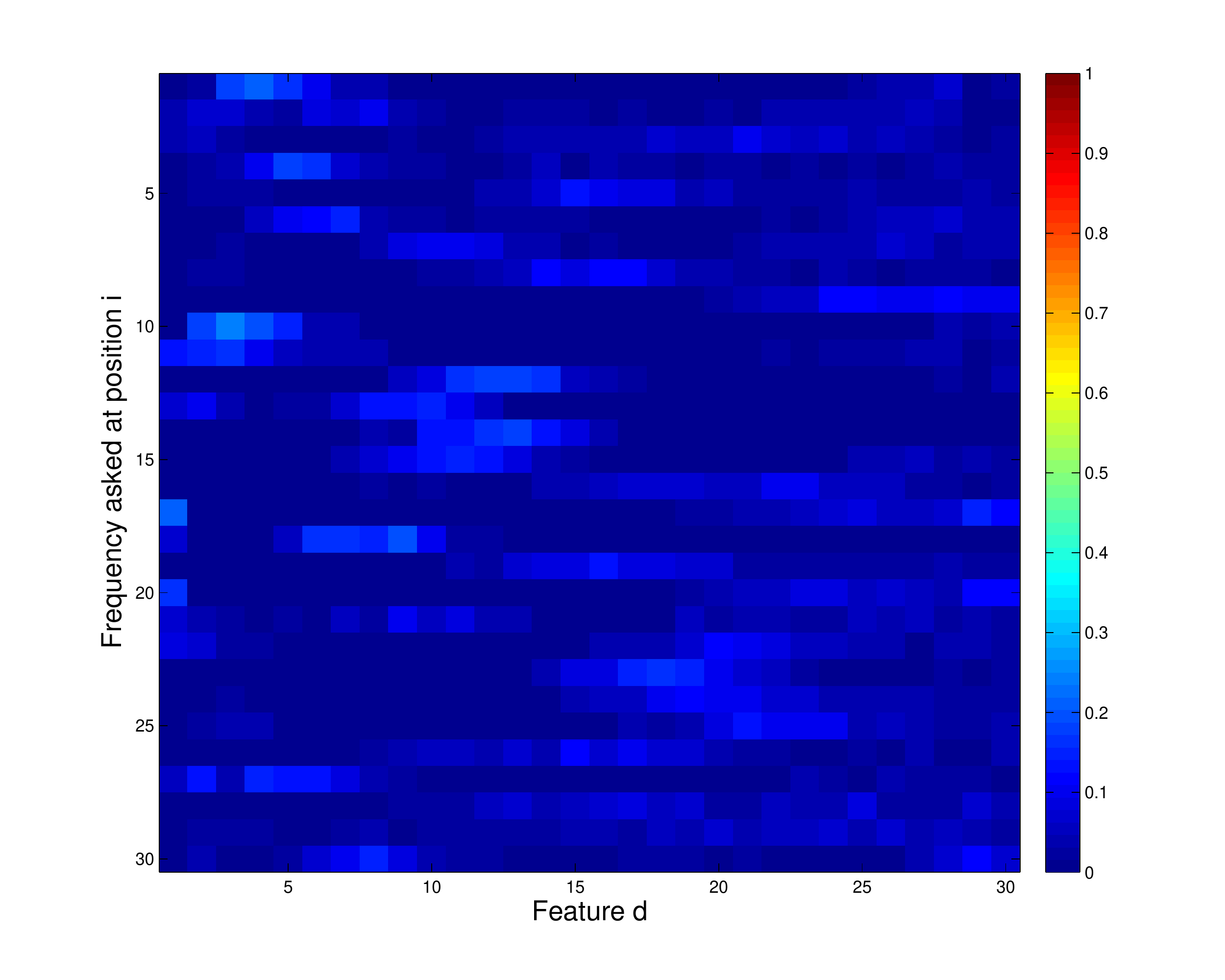}
  \caption{The oracle chose to add features fairly uniformly across test samples, shown here as the frequency each feature was asked in each position of question orders.}
  \label{fig:heatmap-elec}
\end{subfigure} \hfill
\caption{Results from dynamically ordering questions for test-time electricity prediction.}
\label{fig:results}
\end{figure}

\subparagraph{Certainty Metrics}
For certainty, we calculated widths of 90\% prediction intervals as features were answered. Since narrower prediction interval widths correspond to more certain predictions, we expect DQO interval widths to be less than those of the baselines, particularly in the early stages. Figure~\ref{fig:pred-int-elec} plots the \emph{actual} prediction interval widths as questions are asked (calculated with Equation~\ref{eqn:mem-lr-pred-int}, using the true known feature values and imputed values for unknown features), averaged across the test dataset, for the question sets from the seven orderings. The DQO sets result in the narrowest (or near-narrowest) prediction intervals (\emph{i.e.}, most certain predictions), compared to the baselines, with improvements most notable in the first 10 questions answered\textemdash the situation that arises when users do not answer all the questions.

\subparagraph{Error Metrics}
For error, we calculated the absolute value of differences between the midpoint of the 90\% prediction interval and truth as questions are answered, plotted in Figure~\ref{fig:ydiff-elec}. Because this metric compares a point prediction to the true value, error is still incurred for prediction intervals that include the true value (as 90\% of them will, by construction), when the true value is not the exact midpoint of the range. For all orderings, predictions approach the true value as questions are answered. Once about 10 questions have been asked, \emph{DQO with cost} reaches similar performance as \emph{DQO without cost} and the fixed-order baselines (\emph{Fixed decreasing} and \emph{Fixed selection}).

\subparagraph{Cost Metrics}
Progressive total feature costs as features are asked and their true values are used in the models are plotted in Figure~\ref{fig:prog-cost-elec}. Cumulative feature costs are lower for orderings that penalize feature cost (\emph{DQO, Oracle with cost}), with cost decreasing as the penalty on cost $\lambda$ increases. The other orderings have similar cost trajectories to each other.

\subparagraph{}
Overall, these metrics show that our test-time DQO approach quickly achieves accurate, confident predictions: by asking around 10 questions, DQO (with and without cost) reaches similar accuracy as the fixed-order baselines, but the sequential predictions by the fixed orderings are less confident than DQO until about 20 questions have been asked.

\begin{table}[h]
\caption{Areas under the curve for the certainty, error, and cost metrics from various methods, for electricity prediction: smaller values mean the algorithm spent less time in high uncertainty, error, and cost.}
\label{tab:results}
\begin{center}
\begin{tabularx}{\linewidth}{|l*{3}{|>{\centering\arraybackslash}X}|}
\hline
{\bf Method} & \it Interval width  &$|y-\hat{y}|$  &\it Cost \\ \cline{1-4}
DQO without cost     &42.43     &12.06   &1212.91 \\
DQO with cost     &42.44     &12.53   &1120.50 \\
Random     &42.53     &13.18   &1213.30\\
Fixed decreasing     &42.46     &11.85   &1233.50\\
Fixed selection     &42.47     &11.45   &1190.50\\
Oracle without cost     &42.35     &14.62   &1222.91\\
Oracle with cost     &42.39     &13.63    &1120.50 \\
\hline
\end{tabularx}
\end{center}
\end{table}

Figure~\ref{fig:heatmap-elec} shows how frequently the oracle asked each feature in each position across test instances. Most features are chosen fairly uniformly at each position in the question ordering. This indicates that there is no single best order to ask questions across all households, which is why the dynamic question-ordering process is so valuable.

\subsection{Limitations}
Currently, our DQO algorithm assumes that (1) users are able to answer the next question we ask and (2) their answers are accurate. However, situations could arise where these assumptions do not hold. For example, in the utility prediction task, a prospective tenant may be interested in getting personalized energy estimates for a home before visiting\textemdash they could still answer occupant-related features. DQO can be easily extended to this case by offering users a ``don't know" option for answering questions and removing unknown features from consideration in later iterations. Breaking the second assumption, that user answers are accurate, would allow people to give estimates for features (\emph{e.g.}, refrigerator size by looking at pictures in the rental listing). Incorporating this element into DQO would require a way to estimate error associated with user-provided feature estimates.

Furthermore, we have not yet tested the sequential question-asking and prediction-providing process of DQO with human users. We hypothesize that giving users estimates from partial information will motivate them to continue answering questions to receive more accurate personalized predictions. On the other hand, once the prediction interval width is small enough or stable enough, users may no longer see the value in continuing to answer questions and will stop.

\subsection{Conclusion}
Providing personalized energy estimates to prospective tenants with limited, costly information is a challenge. Our solution uses an established dataset to build cost-effective predictive models and, at test time, dynamically orders questions for each user. At test time, when we want to make a personalized estimate for a new renter-home pair, we present a cost-effective way to choose questions to ask a user about their habits and a rental unit, based on which feature's inclusion would most improve the certainty of our prediction, given the information we already know. Our experiments show that, for predicting electricity and natural gas consumption, we achieve prediction performance that is equally accurate to, but more certain than, two fixed-order baselines by asking users only 21\% of features (26\% of the cost of the full-feature model). 

\section{General Framework for Dynamic Question Ordering in Online Surveys}
\label{sec:gen-framework}
The dynamic question-ordering procedure defined in \textproc{DQO\_all} (Algorithm~\ref{alg:dqo-all}) has two components at its core that can be generalized to an online survey. First, there is a way to impute values for unanswered questions. Second, there is a way to select an optimal next question to ask. In this section we expand the methods used to achieve these goals in the previous section and discuss how other components can be used in similar ways when applying DQO to other prediction problems and to survey collection.

\subsection{Imputing Unknown Feature Values}
To estimate answers to yet-unasked questions, the DQO approach presented in Section~\ref{sec:recs-dqo} used $k$ nearest neighbors in \textproc{estimate\_features} (Algorithm~\ref{alg:knn-update}). The appeal of $k$NN here is that, as a nonparametric prediction method, it can estimate values for all unknown features from any subset of known features, simply by finding the $k$ nearest neighbors to the query point restricted to the dimensions currently known (\ie the questions already answered). Using a parametric prediction method, such as linear regression or naive Bayes, to estimate unknown features is challenging, because it would require a model for predicting every feature on every potential subset of features, since questions are not asked in a fixed order. While feasible for relatively short questionnaires, learning an exponential number of predictive models for each question in a survey, from all potential sets of previous questions, is definitely impractical.

\subsection{Selecting an Optimal Next Question to Ask}
To choose an optimal next question to ask, DQO for personalized energy estimates minimized the expected prediction uncertainty from including each potential new question (calculated as the expected width of the prediction interval, in \textproc{Expected\_interval\_width} (Algorithm~\ref{alg:e-pred-int})), penalized by question cost. We used prediction interval width as a measure of uncertainty for regression, but other measures of prediction uncertainty could be used instead. 

Other criteria for question-selection include maximizing information gain from choosing a next question, maximizing an expected response probability, or minimizing an expected breakoff probability (both calculated from items in the survey, or from paradata collected during the survey). Alternatively, item-specific response and breakoff rates could inform the cost of each question.

This method can be generalized to order \emph{modules} of related questions, rather than individual questions. Reasons to present questions in modules rather than purely sequentially include (1) presenting related items in a group can reduce the \emph{cognitive burden} required of a respondent to answer the group (\eg if a set of questions asks the respondent about various aspects of their commute, as the American Community Survey does, it will be easier for the respondent to answer those commute-related questions as a unit rather than scattered throughout the entire questionnaire)~\cite{tourangeau1984cognitive} and (2) imposing a standard order on certain questions that are susceptible to \emph{order effects}~\cite{sudman1996thinking} can ensure that all participants understand and answer questions in the same way, even when question order is determined dynamically.

\section{Other Potential Survey Applications}
\label{sec:other-apps}
In this section, we describe other surveys in which dynamic question ordering could be beneficial, with elaboration on what particular problems DQO could solve. Some of these surveys, such as the Current Population Survey (CPS) and the National Crime Victimization Survey (NCVS), have particular prediction problems as goals (identifying if the respondent is unemployed for the CPS or classifying incidents of victimization for the NCVS) and can therefore directly incorporate a prediction-motivated approach to DQO, like the one presented in Section~\ref{sec:recs-dqo}. Other surveys, like the American Community Survey and National Health Interview Survey, are more focused on the broad goal of collecting information on a large population, rather than making specific predictions for individual respondents. Dynamic question ordering in these surveys would then need to focus on maximizing respondent engagement, calculated from previously-provided answers and paradata collected during the survey-taking. National surveys of these sorts are complicated to deal with, due to complex sampling requirements (such as oversampling certain populations, as the NHIS does) and multi-purpose goals (such as adding supplemental modules to core surveys, as happens in the CPS and NHIS).

\subsection{American Community Survey}
For the mandatory American Community Survey (ACS), the goal is to gather complete statistics on the U.S. population, and follow-up with nonrespondents is expensive. Each year 3.54 million households receive mailed surveys to answer anywhere between 77 and 347 questions, depending on the number of household occupants~\cite{census2016acs}. The survey takes, on average, 40 minutes to complete and 54\% of homes return theirs~\cite{census2014acs}. The Census Bureau calls nonrespondents for telephone interviews and then samples nonrespondents for home interviews. Each in-person case takes 134 minutes; in 2012 this amounted to 129,000 person-hours per month~\cite{griffin2014reducing}. In addition to being expensive, in-person interviews can also bias survey results due to higher weights assigned to those subsampled respondents~\cite{census2014acs}. The Census Bureau tested shifting the mail survey online and found similar data quality for internet and mail return~\cite{horwitz2012data}. Furthermore, while overall response rates were similar, online surveys had higher item response rates for earlier questions and more blank responses for later questions than paper surveys~\cite{horwitz2012data}. Dynamically ordering survey questions in the online form could ensure that even if households do not complete the survey, they answer the most informative questions before breaking off.

The online mode for the ACS also collects paradata as respondents complete the survey. These paradata include clicked links (including navigation buttons, responses, help buttons), timestamps, field values, errors, invalid logins, timeouts, logouts \cite{horwitz2012data}. Such paradata could be used to model user engagement, understanding, and willingness to respond, as another component for dynamic question ordering to increase response rate.

\subsection{Current Population Survey}
The Current Population Survey (CPS) is a monthly survey of 60,000 households across the United States, jointly sponsored by the U.S. Census Bureau and Bureau of Labor Statistics~\cite{census2006cps}. Selected households are in the survey for four consecutive months, out of the survey for eight months, and then back in the survey for four more months. Originally conducted as an in-person paper survey, the CPS first introduced computer-assisted telephone interviewing (CATI) in 1987 and computer-assisted personal interviewing (CAPI) in 1994; currently, the CPS is completely computerized and conducted via CATI and CAPI. These computerized modes would allow for dynamic question ordering as respondents answer survey questions.

The chief purpose of the CPS is to estimate the United States unemployment rate for the past month, and, consequently, the majority of the official survey is devoted to this task. Respondents answer a battery of questions related to their work status in the past week to determine if they were employed, unemployed, or not in the labor force. There are over 200 questions in the labor force portion of the items; not all of these questions apply to every household, so the current version of the CPS uses predefined skip patterns to avoid asking irrelevant questions. Augmenting the rule-based skip patterns with dynamic question ordering derived from statistical properties of the respondent could further lower respondent burden and response time.

Various survey sponsors add supplemental question modules to the CPS (\eg the Tobacco Use Supplement, sponsored by the National Cancer Institute), which may also benefit from dynamic question ordering. The number of supplements is heavily restricted, due to not wanting to overburden respondents with too many questions and detract from the main purpose of the survey\textemdash estimating employment rate~\cite{census2006cps}. Using dynamic question ordering within or between modules could effectively select items to ask of populations of interest, thereby reducing the effective number of questions respondents must answer and increasing the potential for supplemental questions on the CPS.


\subsection{National Health Interview Survey}
The largest U.S. health survey, the National Health Interview Survey (NHIS) is administered in person to about 35,000 households, representing 87,500 individuals, throughout each year~\cite{nchs2015nhis}. However, this number can increase or decrease, depending on available funding; for example, the 2014 NHIS covered 44,552 households, representing 112,053 individuals. The main purpose of the NHIS is to collect health-related information on a household, both at the household and family levels and at the individual level for one ``sample child" and one ``sample adult" from each family. 

As a CAPI-conducted survey, the NHIS could feasibly incorporate dynamic question ordering into its interview procedure. Currently, the NHIS uses predefined skip patterns to advance respondents through the survey, but a statistical approach to question order could enhance the survey experience. Like the CPS, the NHIS also has supplements to the main survey sponsored by other agencies. Supplemental questions are often asked in their own modules but are occasionally interspersed into the NHIS Core.

The structure of the NHIS designates one person from a household as the ``household respondent" who provides information for all members in the household (even for multi-family households). This type of proxy reporting is more likely to have errors than self-reports~\cite{sudman1996thinking}, and so a dynamic question-ordering procedure would need to consider the impact of uncertain provided values when choosing which question to ask next.

The NHIS oversamples underrepresented populations, like black, Hispanic, and Asian people (especially when they are at least 65 years old), to obtain more precise estimates for these populations~\cite{nchs2015nhis}. With this goal in mind, a dynamic-question ordering procedure for the NHIS could also take into account the likely accuracy of imputed values for questions that are not yet asked, with thresholds for allowable imputation error. Such thresholds could be population-specific, with minorities' having much lower allowable error thresholds, to ensure that more complete data are collected from members of these populations.

\subsection{National Crime Victimization Survey}
Every year the U.S. Census Bureau, on behalf of the Bureau of Justice Statistics, administers the National Crime Victimization Survey (NCVS) to 90,000 households (160,000 individuals) in the United States~\cite{bjs2014ncvs}. Once selected for the survey, a household's occupants age 12 and older are interviewed every six months over three years, for a total of seven interviews. In interviews, respondents report victimizations, both reported to police and unreported, that they experienced in the previous six months. The interview is always conducted in a computerized mode (CAPI or CATI), with the first interview in person; since 2006 the NCVS has been administered via CAPI, so dynamic question ordering is possible in this computerized setting.

The NCVS collects detailed information about each incident reported by a respondent, to classify incidents into fine-grained categories of crime (\eg ``Robbery -- completed without injury," ``Robbery -- attempted with injury"). The current NCVS design asks a respondent a set of questions regarding each incident they report and uses answers to these questions to classify the crime, rather than directly asking respondents for the crime category~\cite{bjs2014ncvs}. As such, this survey has a per-individual prediction problem at its core (labeling an incident as a type of crime), just like the personalized energy estimate example with RECS presented in Section~\ref{sec:recs-dqo}, and could benefit from a similar DQO process, except for classification rather than regression. 

Dynamic question ordering could further benefit the NCVS because, especially as households complete the survey multiple times, respondents recognize that reporting an incident results in an extended set of questions to answer. This full questioning takes place for each individual report, including repeat victimizations (\eg domestic violence). To speed up the interview, participants are likely to underreport incidents of victimization, to avoid lengthy subsets of questions for each report. By reducing the number of questions necessary to categorize each incident and using previously provided information to further help in question ordering for repeat incidents, dynamic question ordering could reduce the number of questions in the entire survey, making it much less burdensome for respondents to provide complete reports.

\subsection{Survey of Income and Program Participation}
Conducted by the U.S. Census Bureau, the Survey of Income and Program Participation (SIPP) collects data on income, employment, and social program participation and eligibility from households~\cite{census2014sipp}. The SIPP is designed as a longitudinal national panel survey, where each panel is a representative sample of 14,000 to 52,000 households, contacted yearly for three to five consecutive years. Each household interview is conducted in person, via CAPI, and aims to get self-reports from all household members at least 15 years old. In addition to demographic information, interviews ask respondents for their participation in various social programs, financial situation, and employment status, in the previous calendar year.

The chief goal of SIPP is to understand household program eligibility and participation and to assess the effectiveness of social programs like Supplemental Security Income, Supplemental Nutrition Assistance Program, Temporary Assistance for Needy Families, and Medicaid. Using participation in each program of interest as the prediction of interest could guide a DQO approach as illustrated in Section~\ref{sec:recs-dqo} on RECS.

\section{Future Work}
\label{sec:future}
Although the supposed neutrality of the survey as an impartial data collection tool means that all respondents have the same (or very similar) survey experiences, this rigid structure can also hinder the natural flow of information that occurs in a conversation~\cite{suchman1990interactional}. Often for a participant, a particular event influences their answers for multiple questions; however, unless a direct question about this event appears in the survey, they have to answer many repetitive questions that could have been avoided in a conversation. Learning a latent structure of participants' answers in a survey could be a step toward uncovering these hidden events that determine the answers to multiple questions, and DQO could use this knowledge to guide question selection as well.  

As we mentioned at the outset, the cognitive aspects of survey methodology movement that originated in the 1980s \cite{jabine1984CASM,tanur1992questions} raised issues with the traditional approach to survey questionnaire design, which keeps order fixed for all respondents and which measures the same quantities at different points in time.  The need to reduce respondent burden and to keep respondents engaged in online surveys is raising a complementary set of issues that are now being addressed under the rubric of adaptive survey design.  These two perspectives do need to be reconciled in some fashion.

In this paper, we considered the prediction-focused implementation of DQO as a special case of the more general survey-taking setting. However, given typical survey respondents' disengagement from surveys and declining survey response rates, maybe a new paradigm of survey collection, in which respondents get something useful to them out of answering a survey, could motivate participants to provide complete and accurate responses. Commercial surveys often pay respondents, but compensation does not necessarily ensure thoughtful responses\textemdash participants still exhibit satisficing behavior in paid surveys (\eg~\cite{barge2012using,kapelner2010preventing}). Incentivizing respondents with something dependent on the quality of their answers, like a personalized prediction or calculation, can motivate them to provide data that accurately reflect their situations. However, one clear downside to this approach is that giving respondents information that comes from the survey they are currently answering contaminates their response. For example, suppose that a person is answering questions about their energy-using habits to receive a personalized energy estimate, as in Section~\ref{sec:recs-dqo}. Their current estimate for natural gas consumption is higher than they would like, and the next question asks for their preferred temperature in the winter. Because they do not want their estimate for natural gas usage to climb even higher, the respondent gives an optimistically-low value for preferred temperature. The uncertainty associated with these predictions can also influence a user's decision to continue answering questions: once a participant feels that their given prediction is certain enough, they may stop answering questions. Depending on the purpose of the survey (namely, whether its chief goal is to provide information to or to collect information from the respondent), this type of breakoff may or may not be bad.

\section{Conclusion}
\label{sec:concl}
Dynamic question ordering\textemdash\ie choosing which question to ask a survey respondent next, depending on their answers to previous questions\textemdash can improve survey quality in two key ways. First, giving participants personalized question orders can engage them and motivate them to complete the survey. Second, eliciting the most relevant information for a particular respondent upfront can improve the quality of imputations for unanswered questions if the respondent breaks off before completing the questionnaire. For some surveys, the goal is only to estimate a value for each respondent; in this case, it is not even necessary for the participant to answer all questions\textemdash it is sufficient for them to answer a subset that will ensure a confident prediction. 

We present a general framework for dynamic question ordering in online surveys that sequentially considers which question to ask a respondent next, based on their previous answers, trading off the expected utility of having an answer to that question with the cost of asking that question. The definition of ``utility" for an answer depends on the survey and its purpose; examples include information gain, response probability, (negative) breakoff probability, or certainty of the subsequent prediction. Similarly, the definition of question ``cost" also depends on the survey; examples include difficulty to the user of answering the question, (negative) likelihood of answering (since respondents may be reluctant to respond to sensitive questions, even if they are easy to answer), or breakoff rates of individual questions.

We illustrated an example of this DQO framework for a prediction-oriented survey\textemdash providing prospective tenants with personalized energy estimates in potential homes. In this application, we found that asking users, on average, 21\% of 30 questions could provide certain and accurate predictions at only 26\% of the cost of the full-feature model, and that there was no fixed order of questions that was optimal across all users. Then, we discussed ways that dynamic question ordering could improve quality in computerized national surveys, focusing on unique aspects of each survey that DQO must take into account.

As more surveys move online or to computerized modes, dynamic question ordering can improve survey results at scale and at low cost to the data collectors. DQO trades off the utility from having an answer to a question with its cost and sequentially requests feature values in order to make useful, confident predictions and gather survey data with the resources users are willing and able to provide.

\bibliographystyle{plain}
\bibliography{dqoos}
\end{document}